\documentclass[reprint,amssymb,amsmath,aip,cha,twocolumn]{revtex4-1}
\usepackage{graphicx}
\usepackage[caption = false]{subfig}
\usepackage{color}
\usepackage{epsfig}
\usepackage{makeidx}
\usepackage{ifpdf}
\usepackage{url}%
\usepackage{cleveref}
\usepackage{float}
\usepackage{bm}
\usepackage[colorlinks=true,linkcolor=blue]{hyperref}%
\expandafter\ifx\csname package@font\endcsname\relax\else
 \expandafter\expandafter
 \expandafter\usepackage
 \usepackage{varioref}
 \expandafter\expandafter
 \expandafter{\csname package@font\endcsname}%
\fi
\begin{document}
\title{Effect of polarization force on the Mach cones in a complex plasma}
\author{P. Bandyopadhyay$^{1,2,}$\renewcommand{\thefootnote}{\alph{footnote}} \footnote{Electronic mail: pintu@mpe.mpg.de}, K. Jiang$^{1,3}$, R. Dey$^4$, G. E. Morfill$^1$}
\affiliation{$^1$ Max-Planck Institut f\"{u}r Extraterrestrische Physik, D-85741, Garching, Germany\\
$^2$ Institute for Plasma Research, Bhat, Gandhinagar-382428, India\\
$^3$ Infineon Technologies AG, Am Campeon 1-12, D-85579 Neubiberg, Germany\\
\mbox{$^4$ Max-Planck Institut f\"{u}r Plasma Physik, Boltzmannstr. 2, 85748, Garching, Germany}}
%\ead{pintu@mpe.mpg.de}
%\date{\today}
%#####################################################################################
\begin{abstract}
We report the modifications of compressional Mach cone propagation characteristics due to \textcolor{black}{the} polarization force acting on micron size dust particles embedded in a non-uniform plasma. We solve the hydrodynamic fluid equations for highly charged dust particles to investigate the Mach cone by incorporating the polarization force in the momentum equation and observe the structural change on lateral wakes at different polarization force for a given Mach number and Epstein drag force. We also notice that the maximum amplitude of normalized dust density perturbation decreases with the increase of polarization interaction when the other parameters remain constant. 
\end{abstract}
%\pacs{52.27.Lw, 52.35.Mw, 47.40.-x}
%\submitto{\NJP}
\maketitle
\section{Introduction} \label{sec:intro}
Mach cones are V-shaped disturbances or shock waves produced by a supersonic/subsonic object moving through a medium. Cone-shaped pattern left behind the trails of a duck or boat/ship in water motivates the researchers to explore it in their respective fields. Complex plasma (or dusty plasma)\cite{Ikezi,Thomas,Hayashi,Chu},  \textcolor{black}{which} consists of electrons, ions and micron sized charge dust grains, serves as an excellent medium in which wake structures can be easily excited. Over the past decade, there has been a great deal of interest in understanding the Mach cone structures excited by a moving disturbance in a complex plasma both theoretically \cite{Havnes1,Havnes2,Dubin,Ma,Zhdanov1,Mamun,Hou1,Jiang1,Hou2,Zhdanov2} and experimentally \cite{Samsonov1,Samsonov2,Melzer,Nosenko1,Nosenko2,Jiang2,Mierk}. The existence of Mach cone in complex plasma was first predicted by Havnes \textit{et al.} \cite{Havnes1,Havnes2} and later it was experimentally observed by Samsonov \textit{et al.} \cite{Samsonov1,Samsonov2} and Melzer \textit{et al.} \cite{Melzer} in a 2D plasma crystal. In the above experiments the compressional Mach cones were excited either by a charged particle moving spontaneously beneath the 2D lattice \cite{Samsonov1,Samsonov2} or by radiation pressure from a well focused laser beam scanning across a 2D crystal \cite{Melzer}. Later on, Nosenko \textit{et al.}\cite{Nosenko1,Nosenko2} performed an \textcolor{black}{experiment} on shear driven Mach cone which was composed of \textcolor{black}{a} single cone. Motivated by these experimental observations on different kinds of wake structures, a number of theoretical models have been proposed to explain the experimental findings. Dubin \cite{Dubin} claimed that the Mach cones are the superposition of linear dispersive waves excited by the moving disturbance which was verified by Ma and Bhattacharjee \cite{Ma} by performing Molecular Dynamic (MD) simulations for compressional Mach cones in the far field approximations.\par
An attention has also been paid to study the Mach cone under the micro-gravity($\sim 0g$) conditions \cite{Thomas2} in a three dimensional crystal/fluid. Couple of experimental observations of 3D Mach cone were reported \textcolor{black}{independently} by Jiang \textit{et al.}~\cite{Jiang2} and Mierk \textit{et al.}~\cite{Mierk} under the $0g$ condition on board International Space Station. An observation of single cone Mach structure~\cite{Jiang2} was reported for the first time in a 3D fluid and the results were compared with a hydrodynamic model. Soon after, the speed of sound was measured directly by exciting a double Mach cone structure using a supersonic particle of Mach number $M\lesssim 3$ \cite{Mierk}. However, a theoretical description of shear-wave Mach cone in \textcolor{black}{a} 3D strongly coupled complex plasma has already been proposed by Bose \textit{et al.} \cite{Bose} by developing a generalized hydrodynamic model to describe the formation of single Mach cone \textcolor{black}{structure}.\par 
Recently, a series of theoretical studies \textcolor{black}{has} been performed to investigate the propagation characteristics of linear and non-linear Dust Acoustic Waves (DAWs) in \textcolor{black}{the} presence of polarization force acting on dust grains in a background of non-uniform plasmas \cite{Khrapak, Pintu, Mamun2}. The concept of polarization force and its importance in dusty plasma physics \textcolor{black}{were} first discussed by Hamaguchi and Farouki \cite{Hamaguchi1, Hamaguchi2}. \textcolor{black}{The p}olarization force arises due to any kind of deformation of the Debye sheath around the dust grains in the background of non-uniform plasmas. Non-uniformity in plasmas arises due to non-zero gradient of local electron and ion density and/or their temperatures. According to Hamaguchi and Farouki \cite{Hamaguchi1, Hamaguchi2}, the polarization force acting on the micron size dust particles can be expressed as, $F_p=-Q^2\nabla\lambda_d/8\pi\epsilon_0\lambda_d^2$, where $Q=-Z_de$ is the average charge ($Z_d$ represents the charge number) on each particle and $\lambda_d$ is the Debye radius of plasma which can be defined as $\lambda_d={\lambda_{de}\lambda_{di}}/{\sqrt{\lambda_{di}^2 + \lambda_{de}^2}}$ with $\lambda_{di(e)}$ being the ion (electron) Debye radius and is given by  $\lambda_{di(e)}=(\epsilon_0k_BT_{i(e)}/n_{i(e)}e^2)^{1/2}$, $T_{i(e)}$ and $n_{i(e)}$ are the temperature and density of \textcolor{black}{ions(electrons)}, respectively. The main features of polarization force are a) it always acts opposite to the electrostatic force, b) it is independent of polarity of dust grains\textcolor{black}{,} and c) it arises because of polarization of plasma ions around the negatively charged dust grains. The effect of polarization force was first applied by Khrapak \textit{et al.} \cite{Khrapak} to investigate the propagation characteristic of DAW. They showed \textcolor{black}{that} the wave phase velocity of DAWs decreases with the increase of the strength of polarization force. They also mentioned that the effect of polarization force on dust particles is pronounced for bigger particles. However, there exists a critical dust size beyond which the net force on the dust grains is no longer a restoring force and the dispersion relation  admits a transition from propagating DAWs to periodically growing perturbations. The threshold limit on dust grain size depends on plasma parameters and for a typical gas discharge plasma the threshold limit is $\sim 10~\mu$m. The earlier studies on Dust Acoustic Solitary Waves (DASWs) \cite{Pintu,Mamun2} also showed considerable modifications of propagation characteristics in \textcolor{black}{the} presence of polarization force. \par
In this article, we investigate the propagation characteristics of Mach cone in \textcolor{black}{the} presence of polarization force acting on micron sized charged dust particles in \textcolor{black}{a} background of non-uniform plasma. We construct the \textcolor{black}{expressions} for the perturbed dust density and the velocity vector field by solving the hydrodynamic model for dust fluids assuming electrons and ions \textcolor{black}{to be} Boltzmannian. By performing a detailed study, we conclude \textcolor{black}{that} the polarization force plays an important role to determine the Mach cone structures. We also notice that the amplitude of perturbed density reduces with the increase of polarization force when the other parameters remain constant.\par    
The paper is organized as follows. In the next section we discuss the model equations that include the effects of polarization force. In Section III, we discuss our numerical results. A brief concluding remark will be made in section IV.
\section{Theoretical model}\label{sec:theory}
In the standard fluid model of dusty plasma for studying low frequency phenomena in the regime where the dust dynamics \textcolor{black}{are} important, it is necessary to consider the electrons and ions \textcolor{black}{as} light fluids and can be described by Boltzmann distribution and to use the full set of hydrodynamic equations to describe the dust dynamics. The densities of electrons and ions at temperature $T_e$ and $T_i$ can be written as,
\begin{eqnarray}  
\nonumber
n_e=n_{eo}\exp(e\phi(\mathbf{r},t)/k_BT_e),\label{eqn:e_density}\\
n_i=n_{io}\exp(-\textcolor{black}{e}\phi(\mathbf{r},t)/k_BT_i). \label{eqn:i_density}
\end{eqnarray}
\textcolor{black}{H}ere, $n_{eo} (n_{io})$ is the equilibrium density of electrons (ions) and $\phi(\mathbf{r},t)$ is the electrostatic potential in a Cartesian coordinate system with $\mathbf{r}(x,y,z=0)$ of a 2D system. $e$ and $k_B$ denote the electronic charge and the Boltzmann constant, respectively.\par
The fluid equations \textit{i.e}, the continuity, the momentum and the Poisson equation for the dust component in our system can be written as, respectively,
\begin{eqnarray} 
\frac{\partial n_d(\mathbf{r},t)}{\partial t}+\nabla.[n_d(\mathbf{r},t)\mathbf{v_d}(\mathbf{r},t)]&\hspace*{-1.1in}=0,\label{eqn:con}\\ \nonumber
m_d\frac{\partial \mathbf{v_d}(\mathbf{r},t)}{\partial t}+m_d\mathbf{v_d}\nabla. \mathbf{v_d}(\mathbf{r},t)&=Z_de\nabla \phi(\mathbf{r},t)+ \mathbf{F}_{EP} \\&+ \mathbf{F}_p, \; \text{and} \label{eqn:mom}\\ \nonumber
\nabla^2\phi(\mathbf{r},t)=\frac{e}{\epsilon_0}[Z_dn_d(\mathbf{r},t) + Z_t\delta(\mathbf{r}-&\hspace{-0.7in}\mathbf{u}t)+ n_e(\mathbf{r},t)\\& - n_i(\mathbf{r},t)].
\label{eqn:pos}
\end{eqnarray}
In the above equations, $n_d(\mathbf{r},t)$ and  $\mathbf{v_d}(\mathbf{r},t)$ represent the instantaneous number density and the velocity vector of the dust fluid, respectively. $m_d$ denotes the mass of dust particles. In our calculations systematic or stochastic dust charge variation\textcolor{black}{s} are neglected. \par 
We also include a $\delta$-function in the Poisson equation (Eq.~\ref{eqn:pos}) to incorporate the presence of projectile particle which moves with velocity $\mathbf{u}$ and charge number $Z_t$. \textcolor{black} {The projectile particle in our paper is represented as a point charge and this is only applicable when the particle size is relatively small comparing to the scale of the Mach cone phenomena. Studies with particular interests in finite sized projectiles or evidently bigger than those forming the complex plasma medium are out of the scope of this paper, but could be conducted by means of PIC simulations \cite{Hutchinson, Wojciech}. Generally as it is well known that in gas dynamics, the supersonic object creates a U-shaped Mach cone when it is a sphere, in contrast to a V-shaped Mach cone when it is a needle.} \par
We take into account the effect of electrostatic force, Epstein drag force ($\mathbf{F}_{EP}$) and Polarization force ($\mathbf{F}_p$) in the momentum equation (Eq.~\ref{eqn:mom}) which are acting collectively on each \textcolor{black}{particle}. The Epstein drag force arises due to the collision between the background neutral gas molecules and dust particles, which can be expressed as, \cite{Epstein},
\begin{eqnarray}
\mathbf{F}_{EP}&\hspace{-0.6in}=-\gamma_{EP}m_d\mathbf{v}_d(\mathbf{r},t)&\nonumber\\&=-\delta_{EP}\frac{4\pi}{3}n_nm_nv_na^2\mathbf{v}_d(\mathbf{r},t),
\end{eqnarray}
where, $n_n, v_n, m_n$ are gas density, thermal velocity and mass for neutral gas molecules, respectively. $a$ is the dust particle radius. $\gamma_{EP}$ and $\delta_{EP}$ are the Epstein drag coefficient and a constant which depends on the interaction between the dust particles and gas molecules. \par
\textcolor{black}{As discussed in Sec.~\ref{sec:intro}, when the electrons and ions are considered to be Bolzmannian, the polarization force  can be expressed as \cite{Hamaguchi1,Hamaguchi2},
\begin{eqnarray}
\mathbf{F}_{P}=-\frac{Q^2}{8\pi\epsilon_0}\frac{\nabla\lambda_d}{\lambda_{d}^2}
\label{eqn:pol_force}
\end{eqnarray}
In the above situation when plasma satisfies the quasineutrality condition and there is no temperature gradient in the plasma then $\nabla\lambda_d$ can be expressed as,
\begin{eqnarray}
\nabla\lambda_d=\frac{1}{2k_BT_i}\left(1-\frac{T_i}{T_e}\right)\lambda_deE.
\label{eqn:lambda_d}
\end{eqnarray}
Where $E=-\nabla\phi(\mathbf{r},t)$, denotes the elctrostatic field. Using the expression of $\nabla\lambda_d$ from Eq.~(\ref{eqn:lambda_d}) in Eq.~(\ref{eqn:pol_force}) we get,
\begin{eqnarray}
\mathbf{F_P}&=&Q\frac{1}{16\pi\epsilon_0}|Q|e\left(1-\frac{T_i}{T_e}\right)\frac{\nabla\phi(\mathbf{r},t)}{\lambda_dk_BT_i}\nonumber\\&=&-Z_de\Re\nabla \phi(\mathbf{r},t).
\end{eqnarray}}
Where the coefficient, $\Re (= \frac{1}{16\pi\epsilon _0}|Q|e(1-\frac{T_i}{T_e})/\lambda_d k_B T_i$) determines the strength of plasma-particle polarization interaction \cite{Khrapak}. For typical complex plasmas, $a=1\mu m$, $Q=10^3e, \lambda_d=10^{-4}$ m and $T_i=0.03$ eV, we have $\Re=0.12$. For larger sized dust grains which can retain higher values of $Q$ the value of $\Re$ becomes larger and \textcolor{black}{approaches} unity. \textcolor{black}{However, for} $\Re>1$ the polarization force exceeds the electrostatic force and the net force acting on the particles is no longer a restoring force which results in a growing unstable perturbation.\par
Replacing the expression of $\mathbf{F}_{EP}$ and $\mathbf{F}_{p}$ in the momentum equation (Eq.~\ref{eqn:mom}) we obtain the modified momentum equation as,
\begin{eqnarray}
\frac{\partial \mathbf{v_d}(\mathbf{r},t)}{\partial t} + \mathbf{v_d}\nabla. \mathbf{v_d}(\mathbf{r},t)&=\frac{Z_de}{m_d}(1-\Re)\nabla \phi(\mathbf{r},t) \nonumber\\&-\gamma_{EP}\mathbf{v}_d(\mathbf{r},t).
\label{eqn:mod_mom}
\end{eqnarray} 
In the perturbed situation, assuming a first order approximation, the dynamical variables $n_d(\mathbf{r},t)$, $\mathbf{v_d}(\mathbf{r},t)$, $\phi(\mathbf{r},t)$, $n_e(\mathbf{r},t)$ and $n_i(\mathbf{r},t)$ about the unperturbed states are given by,
\begin{eqnarray}
\nonumber
&n_d(\mathbf{r},t)&=n_{d0}+n_{d1}(\mathbf{r},t),\label{eqn:nd}\\ \nonumber
&\mathbf{v_d}(\mathbf{r},t)&= \mathbf{v_{d1}}(\mathbf{r},t),\label{eqn:vd}\\
&\phi(\mathbf{r},t)&=\phi_1(\mathbf{r},t) ,\label{eqn:phi}\\\nonumber
&n_e(\mathbf{r},t)&=n_{e0}+n_{e0}\left(\frac{e}{k_BT_e}\right)\phi_1(\mathbf{r},t),\label{eqn:ne}\\\nonumber
&n_i(\mathbf{r},t)&=n_{i0}-n_{i0}\left(\frac{e}{k_BT_i}\right)\phi_1(\mathbf{r},t).\label{eqn:ni}
\end{eqnarray}

The equilibrium electron density $n_{e0}$ and ion density $n_{i0}$ are related to the equilibrium dust density $n_{d0}$ and the dust charge number $Z_d$ by the charge neutrality condition,
%$$$$$$$$$$$$$$$$$$$$$$$$$$$$$$$$$$$$$$$
\begin{eqnarray}  
n_{i0}=n_{e0}+n_{d0}Z_d.
\label{eqn:quasi}
\end{eqnarray}
%$$$$$$$$$$$$$$$$$$$$$$$$$$$$$$$$$$$$$$$$
Using equations (\ref{eqn:phi}) and (\ref{eqn:quasi}) in equations (\ref{eqn:con}), (\ref{eqn:pos}) and (\ref{eqn:mod_mom}) we obtain,
\begin{eqnarray}  
\frac{\partial n_{d1}({\bf{r}},t)}{\partial t} + n_{d0}\nabla {.{\bf{v}}_{d1}({\bf{r}},t)}=&\hspace{-1.1in}0,\\
\frac{\partial {\bf{v}}_{d1}({\bf{r}},t)}{\partial t}=\frac{Z_de}{m_d}(1-\Re)\nabla \phi_1&\hspace{-0.1in}({\bf{r}},t)-\gamma_{EP}{\bf{v}}_{d1}({\bf{r}},t),\\
\nabla^2 \phi_1({\bf{r}},t)=4\pi e[Z_dn_{d1}({\bf{r}},t)&\hspace{-0.45in}+Z_t\delta({\bf{r}}-{\bf{u}}t)]\nonumber\\&+\lambda_{d}^{-2}\phi_1({\bf{r}},t).
\end{eqnarray}
Where $\lambda_d=\left[\frac{\epsilon_0 k_B}{e^2}\left(\frac{T_eT_i}{T_in_{e0} + T_en_{i0}}\right)\right]^{1/2}$, is known as dusty plasma Debye length as discussed before.\par
By using a partial Fourier transform with respect to {\bf{r}} and t dependencies
\begin{eqnarray}  
A({\bf{r},t)}=\int{\frac{d{\bf{k}}d\omega}{(2\pi)^4}A({\bf{k}},\omega)e^{i{\bf{k}.r}-i\omega t}},
\end{eqnarray}
we can obtain the following expressions:
\begin{eqnarray}  
n_{d1}(\mathbf{r},t)=&\frac{\beta}{(2\pi)^4}\int{d\mathbf{k}d\omega e^{i({\mathbf{k}.r}-\omega t)}}\nonumber\\& \times \frac{[1-\epsilon(k,\omega)]\delta(\omega-\mathbf{k.u})}{\epsilon(k,\omega)}, \label{eqn:density}\\
\mathbf{v}_{d1}(\mathbf{r},t)=&\frac{\beta}{(2\pi)^4n_{d0}}\int{d{\bf{k}}d\omega e^{i(\mathbf{k.r}-\omega t)}}\nonumber\\&\times\frac{\omega}{k^2}  \frac{[1-\epsilon(k,\omega)]\delta(\omega-\mathbf{k.u})}{\epsilon(k,\omega)}{\bf{k}}\label{eqn:velocity},
\end{eqnarray}
where dielectric function $\epsilon(k,\omega)$ can be obtained by
\begin{eqnarray}  
\epsilon(k,\omega)=1-\frac{\omega_{pd}^2(1-\Re)}{\omega(\omega+i\gamma_{EP})}\left(\frac{k^2\lambda_d^2}{k^2\lambda_d^2+1}\right) \label{eqn:die}
\end{eqnarray}
with the dust plasma frequency $\omega_{pd}=(\frac{n_{d0}Z_d^2e^2}{\epsilon_0m_d})^{1/2}$ and the ratio of charges $\beta$= Z$_t$/Z$_d$. It is clear from the above equations (Eqs.~\ref{eqn:density} and \ref{eqn:velocity}), both the perturbed density and the velocity are stationary fields in the frame of reference of the projectile particle moving with velocity $\mathbf{u}$.\par
The dispersion relation of low frequency dust acoustic wave can be obtained from Eq.~(\ref{eqn:die}) by setting $\epsilon(k,\omega)=0$ \cite{Rosenberg}. The modified dispersion relation in \textcolor{black}{the} presence of polarization force and Epstein drag force can be written as,
\begin{eqnarray} 
k^2=k_d^2\frac{\omega(\omega+i\gamma_{EP})}{\omega_{pd}^2(1-\Re)-\omega(\omega+i\gamma_{EP})}.\label{eqn:disp}
\end{eqnarray}
%#################################################################
%###################################################################
\begin{table}[hb]
\caption{Dusty plasma parameters used for the numerical computations.} \label{tab:tab1}
\begin{center}
\begin{tabular}{ll}
\hline\hline
Test particle charge number~($Z_t)$:&10000\\
Average charge on each particle~($Q)$:& 4007e\\
Dust particle radius~(a):& 4.5~$(\mu$m) \\
Inter-particle distance~(d): &$\sim$ 230~$(\mu$m) \\
Equilibrium ion density~($n_{i0}$): & $10^{14}$~(m$^{-3})$ \\
Ion(electron) Temperature~($T_i(T_e))$:& 0.1(3)~(eV)\\
\hline \hline
\end{tabular}
\end{center}
\end{table}
%#################################################################
%##############################   Figure 1  ###########################
\begin{figure*}[ht]
\centerline{\hbox{\psfig{file=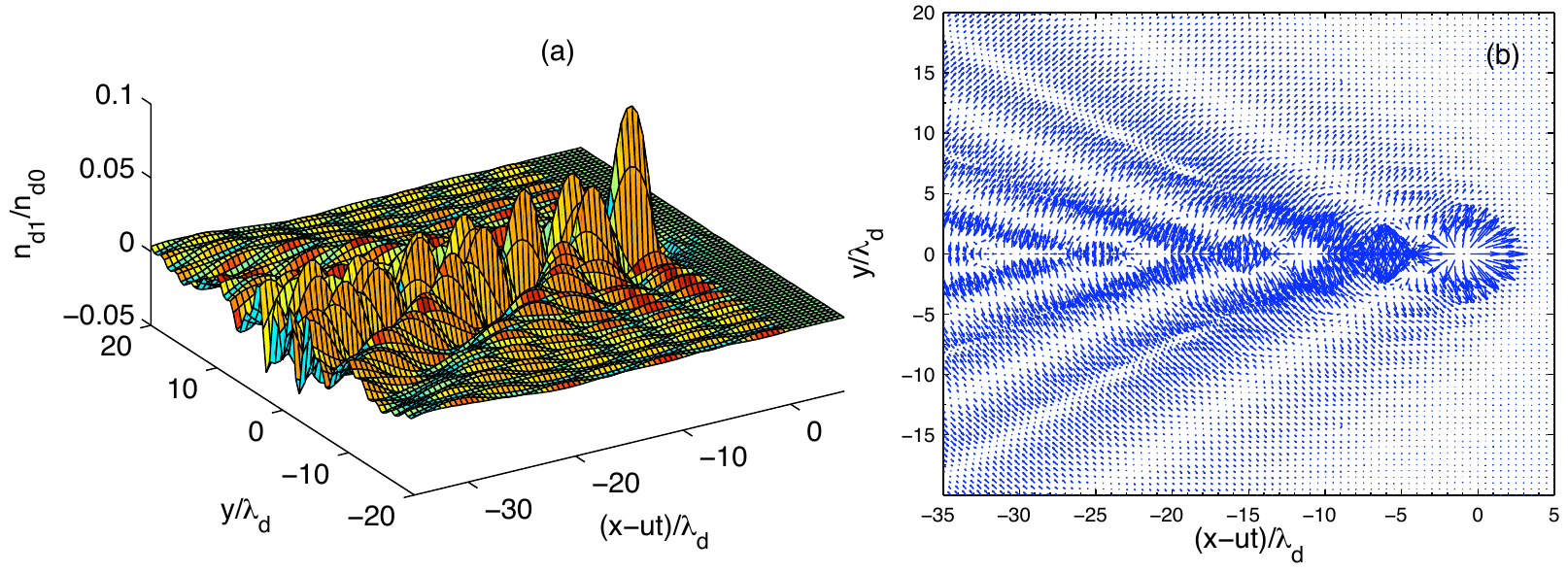,width=.80\textwidth,angle=0}}}
\caption{(Color online). An example of Mach cone structures in presence of polarization force. a) Surface plot of perturbed dust density and b) dust velocity vector map for $M=1.1, \gamma_0=0.01$ and $\Re=0.4$.}
\label{qua_2}
\end{figure*}
%###################################################################################
%##############################   Figure 2  ###########################
\begin{figure*}[ht]
\centerline{\hbox{\psfig{file=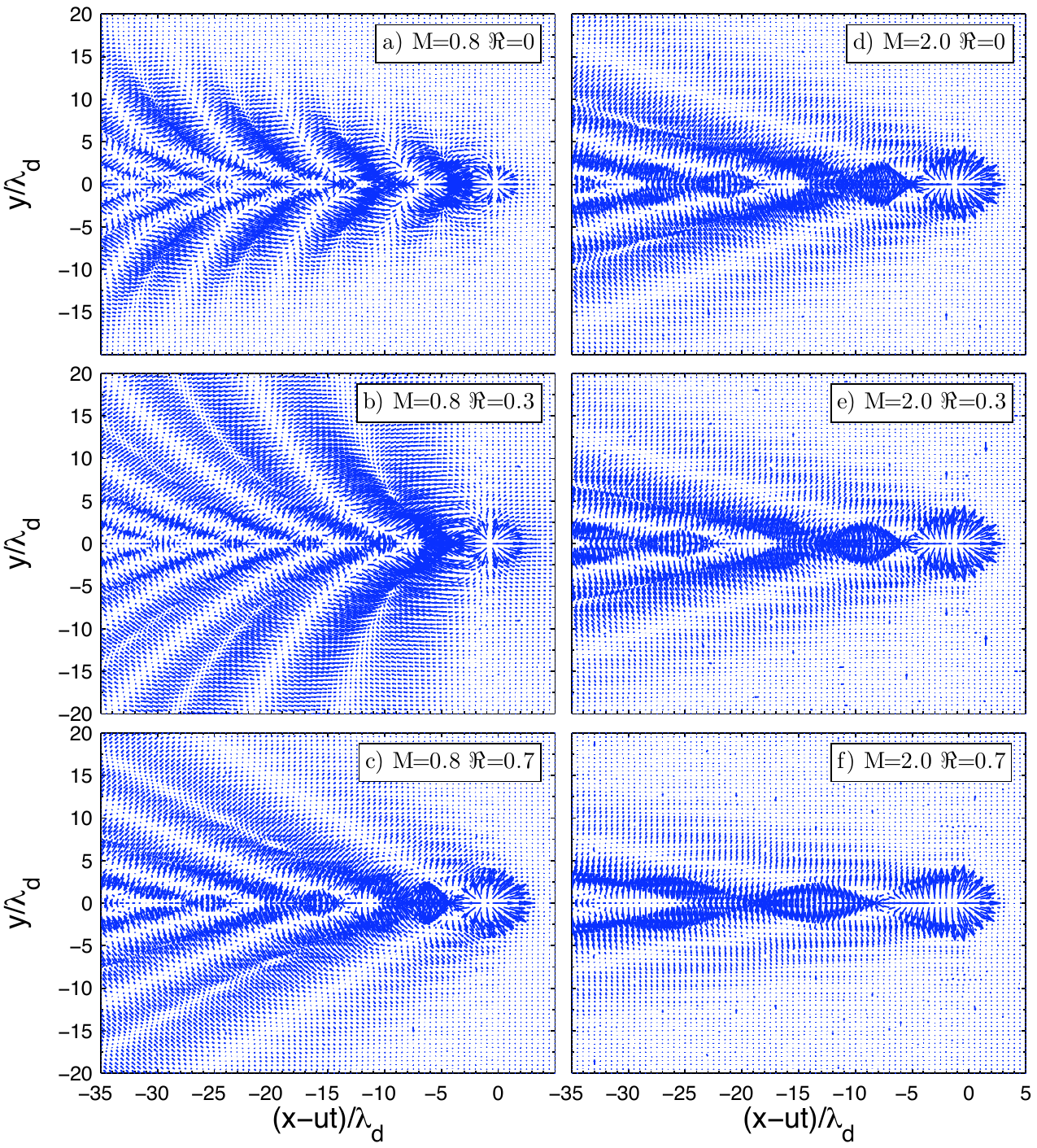,width=0.6\textwidth,angle=0}}}
\caption{(Color online). Map of the particle velocity vector (${\bf{v}}_{d1}$) in case of weak damping $\gamma_{0}=0.01$ for a) $\&$ d) $\Re=0$, b) $\&$ e) $\Re=0.3$ and c) $\&$ f) $\Re=0.7$. The left panel (a--c) is for $M=0.8$ whereas the right panel (d--f) depicts for $M=2.0$.}
\label{qua_2}
\end{figure*}
%###################################################################################
Where we \textcolor{black}{have} used  $k_d=\lambda_d^{-1}$. From this modified dispersion relation (Eq. \ref{eqn:disp}), we can easily re-construct the dispersion relation of Piper and Goree \cite{Piper} by setting $\Re=0$ (absence of polarization force) and the dispersion relation of Khrapak \textit{et al.} \cite{Khrapak} when $\gamma_{EP}=0$ (absence of Epstein drag force).\par
To study the propagation characteristic of Mach cone in detail, we solve the perturbed dust density $n_{d1}$ from Eq.~(\ref{eqn:density}) and the perturbed velocity field vector $\mathbf{v}_{d1}$ from Eq.~(\ref{eqn:velocity}) numerically. In accordance with the recent experiments~\cite{Thomas3}, we choose our basic dusty plasma parameters\textcolor{black}{,} which are tabulated in Table-1. The other parameters \textcolor{black}{ such as,} 2D dust density ($n_{d0}$), mass of MF particle ($m_d$), plasma Debye radius ($\lambda_d$) and dust frequency ($\omega_{pd}$) are estimated with the help of these dusty plasma parameters. It is clear from the above equations (Eqs.~\ref{eqn:density} and \ref{eqn:velocity}), \textcolor{black}{that} the projectile particle velocity ($\mathbf{u}$) plays an important role to determine the wake structure formation which can be expressed in terms of Mach number, $M=u/C_s$~($C_s=\lambda_d\omega_{pd}$ is being the dust acoustic velocity when the particle exerts only the electrostatic force). It is also clear from the above dispersion relation (Eq. \ref{eqn:disp}), \textcolor{black}{that} the dust acoustic velocity changes with the change of polarization force coefficient ($\Re$). It indicates that the modified Mach number does not remain constant with the change of $\Re$. It is also worth mentioning that the normalized damping coefficient, $\gamma_0=\gamma_{EP}/\omega_{pd}$, \textcolor{black}{representing} the background neutral gas pressure, is also an important factor to \textcolor{black}{determine} the Mach cone structures. Although in some cases, we assume \textcolor{black}{that} the Mach number (M) is a constant quantity\textcolor{black}{,} but in reality the Mach number changes with the change of $\Re$ and $\gamma_0$.\par 
%##############################   Figure 3  ###########################
\begin{figure*}[ht]
\centerline{\hbox{\psfig{file=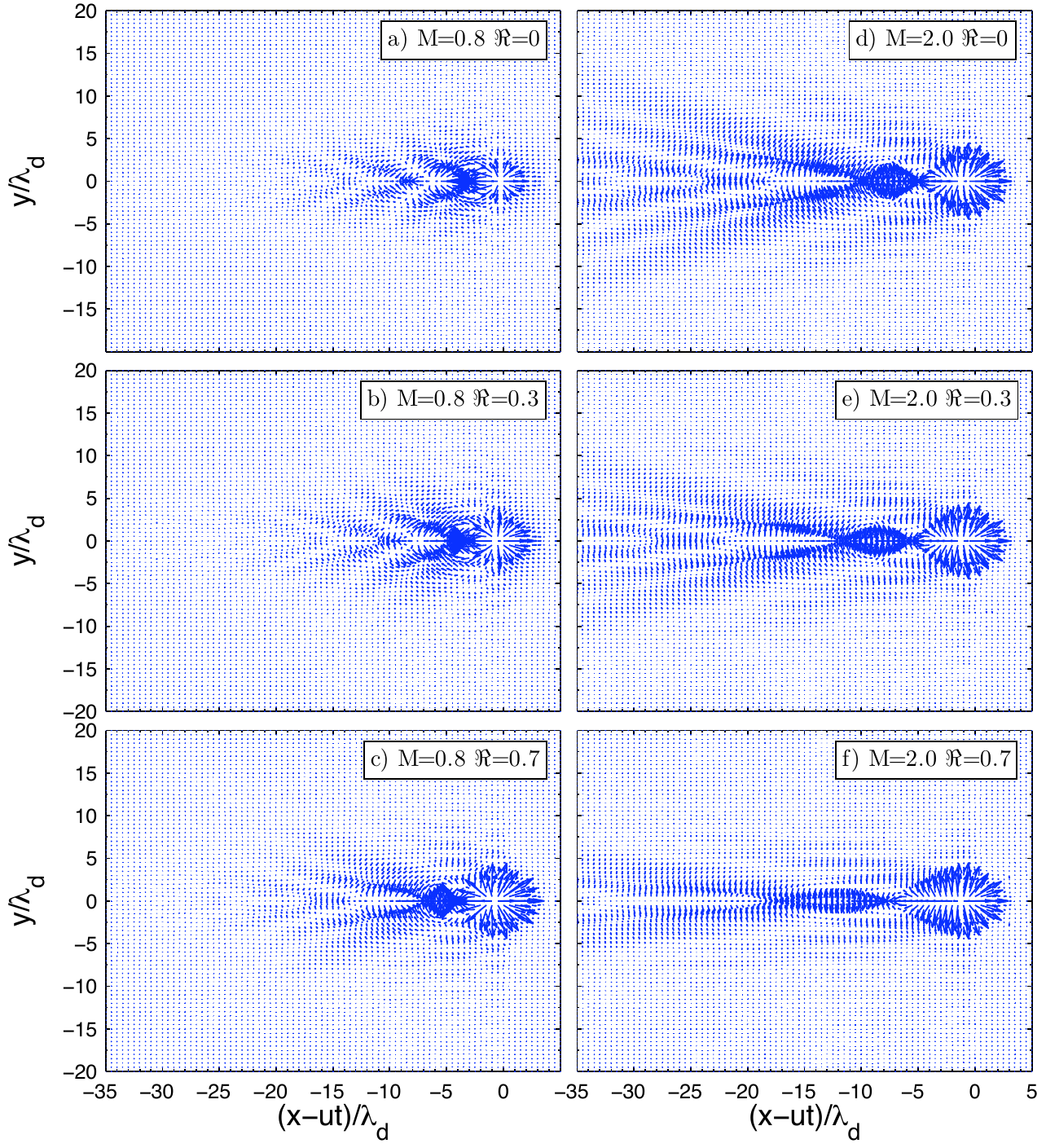,width=0.6\textwidth,angle=0}}}
\caption{(Color online). Map of the particle velocity vector (${\bf{v}}_{d1}$) in case of strong damping $\gamma_{0}=0.2$ for a) $\&$ d) $\Re=0$, b) $\&$ e) $\Re=0.3$ and c) $\&$ f) $\Re=0.7$. The left panel (a--c) is for $M=0.8$ whereas the right panel (d--f) depicts for $M=2.0$.}
\label{qua_2}
\end{figure*}
%###################################################################################

\section{Results and Discussions}\label{sec:results}
Fig.~1(a) and (b) illustrate the typical normalized perturbed density ($n_{d1}/n_{d0}$) and velocity vector map of a lateral wake structure for a given Mach number ($M=1.1$), Epstein drag coefficient ($\gamma_0=0.01$) and polarization force coefficient ($\Re=0.4$). V-shaped Mach cone with multiple wake structures can be clearly seen in both these figures. These patterns are caused by constructive and destructive interference of waves excited by the projectile particle, and their structures are determined by the wave dispersion properties of the medium as well as the disturbance-medium interaction. It also can be seen from the velocity field map (see Fig.~1(b)) that the structures \textcolor{black}{consist} of multiple cones, with the outermost being \textcolor{black}{the} most prominent. According to the earlier investigations of Mach cone in complex plasmas \cite{Dubin}, \textcolor{black}{these} multiple structures arise due to strongly dispersive nature of the dust acoustic waves. This figure also indicates that the direction of the dust particle motion is perpendicular to the cone wings and parallel to the direction of wave propagation. It confirms that the Mach cones are composed of compressional waves. \par
%##############################   Figure 4  ###########################
%%%%%%%%%%%%%%%%%%%%%%%%%%%%%%%%
 \begin{figure}[h]
 \centering
\subfloat{{\includegraphics[scale=0.6]{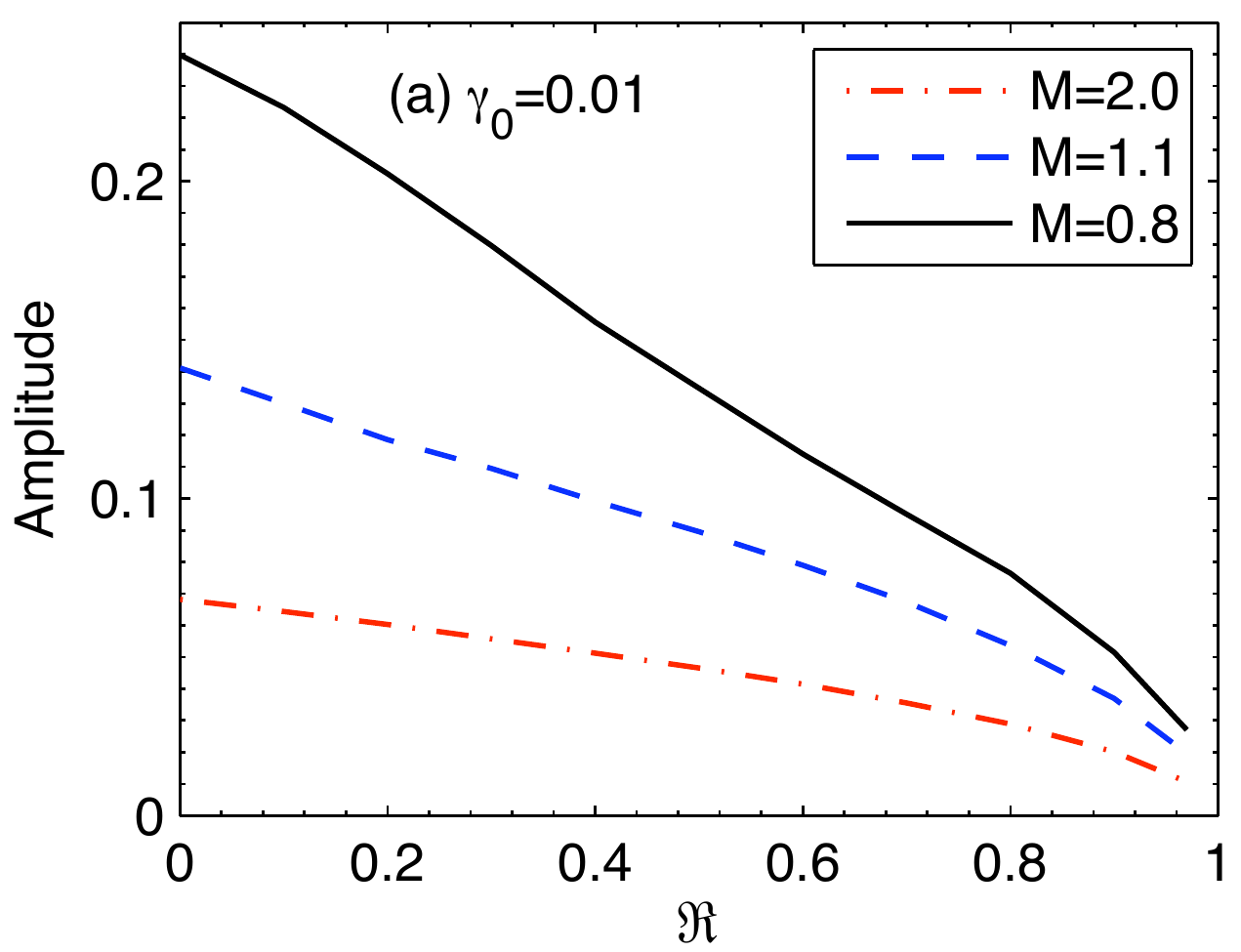} }}%
\vspace*{-0.13in}
  %\qquad
  %\vspace*{-0.in}
    \subfloat{{\includegraphics[scale=0.6]{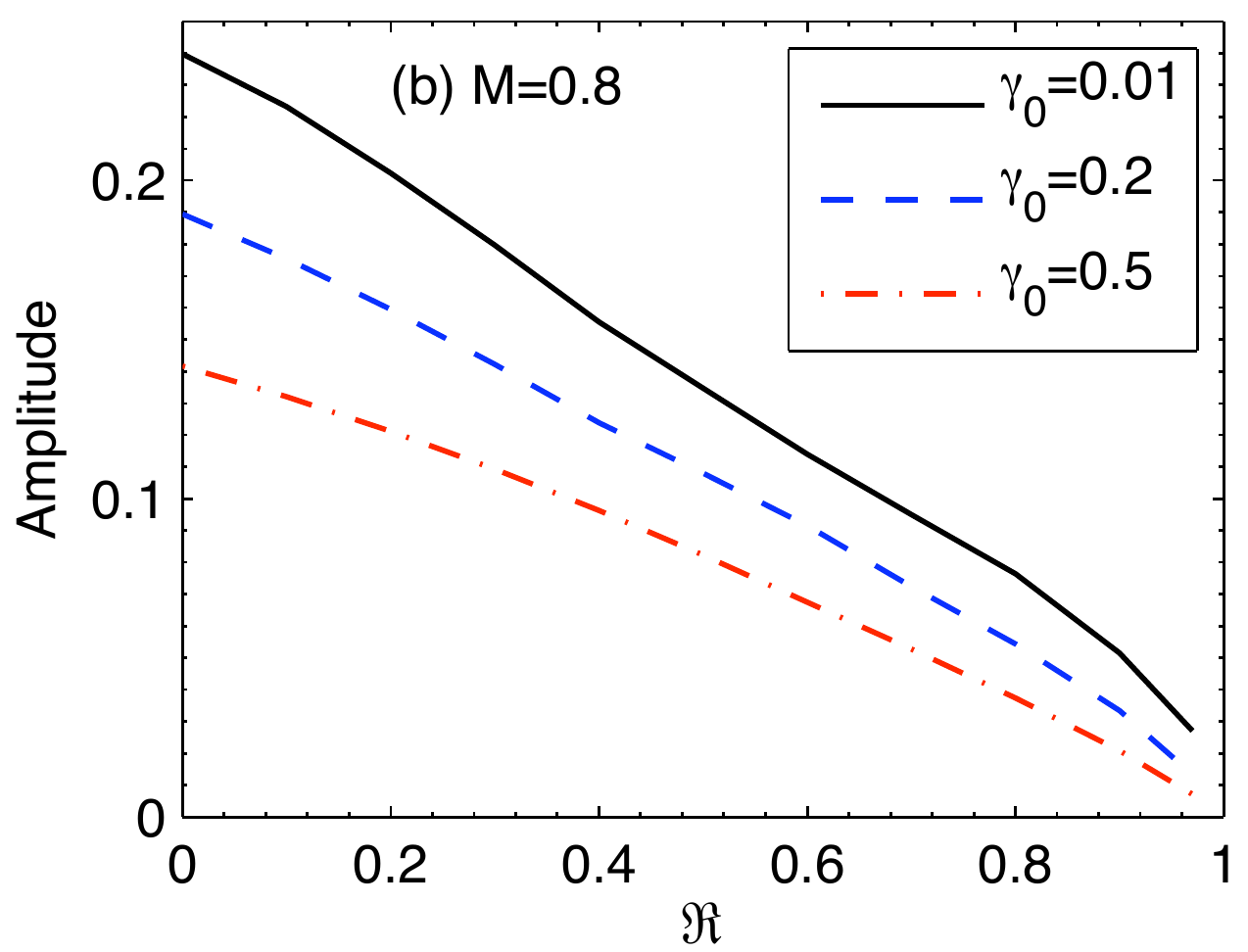} }}%
  \caption{\label{fig:ddw without rod}(a) and (b) Variation of amplitude of normalized perturbed density with $\Re$ for different values of (a) M at constant $\gamma_0$ and (b) $\gamma_0$ at constant M. In (a) solid, dashed, and dashed-dotted lines represent M = 0.8, 1.1, and 2.0, respectively, at $\gamma_0=0.01$. Whereas in (b) solid, dashed, and dashed-dotted lines represent $\gamma_0=0.01$, 0.2, and 0.5, respectively, at M=0.8.}
  \end{figure}
 %%%%%%%%%%%%%%%%%%%%%%%%%%%%%%%%%###################################################################################
To study qualitatively the Mach cone structure for different values of projectile particle velocity in terms of Mach number and the polarization force, we first consider the case when the influence of damping is \textcolor{black}{weak} ($\gamma_0=0.01$). The left panel (a--c) of Fig.~2 shows the velocity field map of Mach cone for a constant value of Mach number ($M=0.8$) at different values of polarization force coefficients, a) $\Re=0$ (absence of polarization effect), b) $\Re=0.3$ and c) $\Re=0.7$. \textcolor{black}{The} right panel (d--f) of Fig.~2 corresponds the same for $M=2.0$ with the same values of $\Re$. It is clear from the left panel of the figure (Fig.~2(a-c)), \textcolor{black}{that} there is a prominent change in the structural properties of the Mach cone when $\Re$ increases at a constant value of $M$ and $\gamma_0$. The number of multiple wake structures reduces while the opening angle decreases with the increase of $\Re$. The changes of opening angle with $\Re$ at constant $M$ can be explained from the dispersion relation (Eq.~\ref{eqn:disp}). With the increase of $\Re$ the velocity of the DAWs decreases resulting in an increase of the modified Mach number. From the Mach-cone-angle relation, the opening angle reduces with the increase of this modified Mach number. It is also noticed \textcolor{black}{that} the  curved nature of the wings of Mach cones \textcolor{black}{becomes} less significant as $\Re$ increases. The bending structures of Mach cones \textcolor{black}{were} also predicted by Zhdanov \textit{et al.} \cite{Zhdanov2}. In their theoretical studies\textcolor{black}{,} they showed that either the time history of changes in the medium or the density gradient in the medium (or both) plays an important role to produce \textcolor{black}{curved} wings Mach cone structures. Although it is not clear in \textcolor{black}{the} present situation which factor is responsible to form the \textcolor{black}{curved} wings but \textcolor{black}{the} changes of the strength of polarization force certainly \textcolor{black}{change} the inhomogeneity of the background plasma medium. It is also to be noted from the right panel of Fig.~2 \textcolor{black}{that} the opening angle and number of multiple structures \textcolor{black}{reduces} (even to one) due to the increase of Mach number. In case of higher Mach number (M=2), the \textcolor{black}{curved} nature of the wings in the Mach cone of velocity map also disappears. \par 
In order to investigate the effect of dust-neutral collision frequency, we increase $\gamma_0$ \textcolor{black}{from 0.01} to 0.2 and plot the velocity vector map in Fig.~3 keeping all \textcolor{black}{other} parameters constant. In accordance with earlier observations \cite{Samsonov2, Nosenko1, Hou1}, the wakes are strongly damped and the oscillations in the wake region are smoothed out due to the decrease of the damping length, $l_d=Cs/\gamma_{EP}$. It is also seen from the above figure that the effect of higher dust-neutral collisionality reduces the number of multiple-wakes structures down to two or even one\textcolor{black}{,} irrespective of the Mach number and the strength of polarization force.  \par
Fig.~4 depicts the variation of maximum amplitudes of perturbed dust density in 2D plasma crystal for a set of Mach number and dust-neutral collision rate. Figure~4(a) represents the same when the dust neutral collision rate remains unchanged at $\gamma_0=0.01$ and the Mach number changes from 0.8 to 2. It is clear from this figure that the amplitude decreases with the increase of $\Re$ for a given Mach number. It suggests \textcolor{black}{that} the increase of $\Re$ influences the wake-field oscillations to be damped which shows a similar effect when $\gamma_0$ increases. As we \textcolor{black}{have} discussed earlier (see in Eq.~\ref{eqn:die}) dielectric function $\epsilon(\omega,k)$ increases with the increase of $\Re$ for a given dispersive medium with constant $\lambda_d$ and $\omega_{pd}$. It can be concluded from the expression of $n_{d1}$ (Eq.~\ref{eqn:density}) \textcolor{black}{that} the perturbed density decreases with the increase of strength of polarization force.  For a given value of $\Re$, it should also be noted that the amplitude decreases with the increase of Mach number. Variation of maximum amplitude as a  function of $\Re$ shows similar trend when we change the dust-neutral collision damping rate from 0.01 to 0.5 at a constant value of Mach number (M=0.8) as shown in Fig.~4(b). This figure also suggests that the amplitude of density perturbation decreases with the increase of $\gamma_0$ for a constant $\Re$.  
\section{Conclusion}
\label{sec:conclusion}
We have studied theoretically the modifications arising in the Mach cone structure due to the presence of polarization force acting on the dust grains in an inhomogeneous plasma. Numerical results for the perturbed density and velocity vector in the dust layers exhibit Mach cone with the characteristic oscillatory wake patterns which is also observed in laboratory experiments \textcolor{black}{\cite{Samsonov1,Samsonov2,Melzer,Nosenko1,Nosenko2,Jiang2,Mierk}}. In our analysis, special \textcolor{black}{attention} is paid to the dependencies on the Mach cone structure for a wide range of polarization force at a given Mach number and the dust-neutral damping rate. Quantitatively, we \textcolor{black}{have} noticed that the amplitude of the perturbed dust density decreases with the increase of the polarization force \textcolor{black}{coefficient}. An experimental observation is needed to verify the polarization-induced change in the Mach cone structures.
%+++++++++++++++++++++++++++++++++++++++++
%\section*{References}

\end{document}